\documentclass[a4paper,10pt]{article}

\usepackage{amssymb}
\usepackage{amsmath}
\usepackage[usenames,dvipsnames]{color}
\usepackage{hyperref}
\usepackage[left=1in,right=1in,top=1in,bottom=1in]{geometry}                  

\hypersetup{
    colorlinks=true,         
    linkcolor=blue,          
    citecolor=red,        
    urlcolor=Violet             
}

\newcommand{\mean}[1]{\ensuremath{\lf\langle #1 \rt\rangle }}
\newcommand{\diby}[2]{\ensuremath{\frac{\delta #1}{\delta #2}}}

\def\be{\begin{equation}}
\def\ee{\end{equation}}
\def\bea{\begin{eqnarray}}
\def\eea{\end{eqnarray}}

\def\lf {\ensuremath{\left}}
\def\rt {\ensuremath{\right}}

\title{Frequently Asked Questions about Shape Dynamics}
\author{\bf Henrique Gomes\footnote{\href{mailto:gomes.ha@gmail.com}{gomes.ha@gmail.com}}\\\it University of California at Davis\\ \it One Shields Avenue Davis, CA, 95616, USA \bigskip\\ \bf Tim Koslowski\footnote{\href{mailto:t.a.koslowski@gmail.com}{t.a.koslowski@gmail.com}}
\\\it Perimeter Institute for Theoretical Physics\\\it 31 Caroline Street, Waterloo, Ontario N2L 2Y5, Canada\\
\it and new address \\
\it  Department of Mathematics and Statistics, University of New Brunswick\\ \it Fredericton, NB, E3B 5A3, Canada}

\begin{document}

\maketitle

\begin{abstract}
  Barbour's interpretation of Mach's principle led him to postulate that gravity should be formulated as a dynamical theory of spatial conformal geometry, or in his terminology, ``shapes.'' Recently, it was shown that the dynamics of General Relativity can indeed be formulated as the dynamics of shapes. This new Shape Dynamics theory, unlike earlier proposals by Barbour and his collaborators, implements local spatial conformal invariance as a gauge symmetry that replaces refoliation invariance in General Relativity. It is the purpose of this paper to answer frequent questions about (new) Shape Dynamics, such as its relation to Poincar\'e invariance, General Relativity, Constant Mean (extrinsic) Curvature gauge, earlier Shape Dynamics, and finally the conformal approach to the initial value problem of General Relativity. Some of these relations can be clarified by considering a simple model: free electrodynamics and its dual shift symmetric formulation. This model also serves as an 
example where symmetry trading is used for usual gauge theories. 
\end{abstract}

\section{Introduction}

The new form of Shape Dynamics, which for the rest of the paper we will refer to merely as Shape Dynamics (SD), is a novel formulation of Einstein's equations in which refoliation invariance is replaced with local spatial conformal (Weyl) invariance. It has first  been laid down explicitly in  \cite{ SD:LT, SD1}, where the adoption of the name Shape Dynamics was motivated by Barbour's interpretation of Mach's principle \cite{Barbour:bm_review,  Barbour:CS+V, Barbour:shape_dynamics}, about which we explain more in the appendix (section \ref{sec:Mach}).  Moreover, the technical  construction of Shape Dynamics utilizes many results of the conformal approach to the initial value problem of ADM, initiated by York et al \cite{York1, York2}. One of the purposes of this paper is  precisely to explain the relation, similarities and differences with previous work, in particular with the Arnowitt-Deser-Misner (ADM) formulation in constant mean curvature (CMC) gauge, the York construction,  and Barbour's 
formulation.

To cater to readers with various backgrounds and various interests in SD, we organized the paper into various modules. We start in section \ref{sec:EM} with a toy model that allows us to introduce the ``best matching-'' and ``linking theory-'' constructions underlaying SD, without having to go through the additional technical complications that are present in a gravity theory. The particular toy model is free electromagnetism with its U(1) gauge symmetry. We apply the Stuckelberg mechanism  w.r.t. axial shifts,  to construct a linking gauge theory. Through the usual completion of the symmetry trading mechanism, this then gives rise to  a shift symmetric theory that can be shown to have the same observable algebra and same dynamics as standard electrodynamics. Having this analogy at our disposal, we give a short description of Shape Dynamics in section \ref{sec:SD}. We proceed with addressing a number of frequently asked questions about SD in section \ref{sec:QA}. We organize the answers in three different 
layers of complexity:  first a simple short answer and then 
more detailed answers for the interested reader; thirdly  technical details for these questions can be found in the appendix (alongside background material).

An introduction to SD has to start with the Machian framework that paved the way for the discovery of General Relativity (GR) and also guided the construction of SD. Barbour's interpretation of the Machian principle asserts that space is abstracted from the relation of objects therein, time is abstracted form change and scale is abstracted from local comparisons. In other words: Barbour asserts that there are no background space, time or scale. Nonetheless, GR implements relativity of space and time, but exhibits an explicit local scale. This is directly linked to the fact that GR has two local degrees of freedom. If gravity would exhibit spatial diffeomorphism-, refoliation- and local spatial scale invariance as independent gauge symmetries then the canonical degree of freedom counting (spatial metric - spatial diffeomorphisn - refoliation - spatial conformal = 6-3-1-1 =1) would result in only one gravitational degree of freedom per point. One can thus not expect to have a purely metric formulation 
equivalent to GR that 
implements 
all three invariances as gauge symmetries. This simple argument does however not preclude trading refoliation for spatial conformal invariance; the resulting theory would have two local degrees of freedom and could thus be equivalent to GR. 

Of course, this simple counting argument does not prove the existence of a locally conformally invariant theory that is dynamically equivalent to GR. We now turn to free electrodynamics to see that gauge symmetries can be traded while maintaining exact dynamical equivalence for all possible observables of the theory.

\section{Warm-Up: Free Maxwell Field}\label{sec:EM}

The concept of ``trading of gauge symmetries" is very generic, but not well appreciated in the literature. It is the purpose of this section to show that symmetry trading it is not a specific technicality due to the structure of gravity, but that it can also take place in every-day gauge theories. The probably best known is electromagnetism.

\subsection{Symmetry trading}

Some important features of the relation of GR, ADM in CMC gauge and SD can already be explained through relating free electromagnetism (the analogue of ADM), electromagnetism in axial gauge (the analogue of ADM in CMC) and the dual shift symmetric theory (the analogue of SD).  

We start the construction with free electrodynamics in a flat background metric $g_{ab}(x)=\delta_{ab}$ using the vector potential $A_a(x)$ and the canonically conjugate electric field $E^a(x)$ and imposing the Gauss constraints $G(\Lambda)=\int d^3x \,\Lambda E^a_{,a} \approx 0$ and the Hamiltonian $H=\int d^3x \,\frac 1 2 \left( g_{ab}E^aE^b+g^{ab}B_aB_b\right)$, where we used the magnetic field $B_a(x)=\epsilon_{abc}A_{c,b}(x)$. Note the important fact that unlike GR, the Hamiltonian here is a \emph{true} one, i.e. \emph{not a constraint}. Therefore its Poisson bracket with second class constraints is not required to vanish, but can be instead set to zero by a judicious choice of the Lagrange multipliers of the second class constraints.  This is the only feature of this model that is not analogous to ADM. 

A very common gauge fixing for electrodynamics is axial gauge, i.e. $A_a(x)v^a(x)=0$ for some fixed vector field $v$. This gauge fixes the local degrees of freedom, but a detailed examination of the spatial topology and boundary conditions is necessary to determine whether all global degrees of freedom are gauge fixed\footnote{E.g. for $v=\partial_3$ on a 3-torus one finds that gauge transformations generated by $\Lambda(x_1,x_2)$ are not gauge fixed, so one needs to introduce additional global gauge conditions.}. To keep the presentation simple and generic, we will first discard all global 
gauge conditions and return to global issues at the end of this section. We will also fix $v=\partial_3$ to simplify the notation, although our formalism works for more general $v$.

We will now mirror the construction of Shape Dynamics in the linking theory formalism (for more on linking theories see appendix \ref{sec:LT-ST}). We first perform best matching w.r.t. shifts in $E^3(x)$ (for more on best matching see appendix \ref{sec:BM}). For this we introduce the auxiliary field $\phi(x)$ and its canonically conjugate momentum $\pi_\phi(x)$ and adjoin the constraints $Q(\rho)=\int d^3x \rho \pi_\phi\approx 0$ to the constraint set, so the auxiliary degree of freedom is unphysical and the theory remains to be electrodynamics. We now consider the generating functional 
\begin{equation}
 \Gamma=\int d^3x \left(\tilde A_a(x) (E^a(x)+\delta^a_3 \phi(x))+\tilde \pi_\phi(x)\phi(x)\right),
\end{equation}
which generates the canonical transform from untilded to tilded variables:
\begin{equation}
 \begin{array}{rclcrcl}
  \tilde A_a(x)&=&A_a(x), &&
  \tilde E^a(x)&=&E^a(x)+\delta^a_3 \phi(x),\\
  \tilde \phi(x)&=& \phi(x), &&
  \tilde \pi_\phi(x)&=& \pi_\phi(x)-A_3(x).
 \end{array}
\end{equation}
This canonical transformation changes the trivially extended theory to
\begin{equation}
 \begin{array}{rcl}
   H&=&\int d^3x \,\frac 1 2 \left( \delta_{ab}E^aE^b+\delta^{ab}B_aB_b+(\phi^2+2E^3\phi) \right)\\
   G(\Lambda)&=&\int d^3x \,\Lambda \left(E^a_{,a}+\phi_{,3}\right)\approx 0\\
   Q(\rho)&=&\int d^3x \,\rho \left(\pi_\phi-A_3\right)\approx 0.
 \end{array}
\end{equation}
This is a linking theory: imposing the partial gauge fixing conditions $\phi(x)=0$ leads to the phase space reduction $(\phi,\pi_\phi)\to(0,A_3)$, which eliminates the shift constraints $Q(\rho)$ and turns $G(\Lambda)$ and $H$ into the Gauss constraints and Hamiltonian of free electrodynamics  The Dirac bracket of the phase space reduction turns out to coincide with the Poisson bracket. Hence this gauge fixing returns us to the original electrodynamics system.

The dual shift symmetric theory is obtained by imposing the gauge fixing conditions $\pi_\phi(x)=0$. This condition together with the Gauss constraint forms a second class system, which yields a partial gauge fixing of the Gauss constraint:
$$ \{\pi_\phi, G(\Lambda)\}=\Lambda_{,3}
$$
This second class system can be solved with the phase space reduction
\begin{equation}\label{equ:reduction}
 (\phi(x),\pi(x)) \to \left(\phi_o[E^a; x_1,x_2,x_3)=f(x_1,x_2)-\int^{x_3}ds \,E^a_{,a}(x_1,x_2,s) ,0\right).
\end{equation}
where we have used DeWitt's notation for mixed functional and spatial dependence (which we largely omit from now on to simplify notation) and $f(x_1,x_2)$ is a general function of the two first coordinates. We note that with appropriate boundary conditions \be\label{equ:phi_o}
\phi_o(x)=-E^3(x)+F[E^1,E^2;x)
\ee where 
$$F[E^1,E^2;x_1,x_2,x_3)=-\int^{x_3}ds\left(E^1_{,1}(x_1,x_2,s)+E^2_{,2}(x_1,x_2,s)\right).$$ 
The dual Hamiltonian is thus independent of $E^3(x)$ and hence invariant under shifts in $E^3(x)$:
\begin{equation}
 H_{dual}=\int d^3x\,\frac 1 2 \left((E^1)^2+(E^2)^2+(\vec B)^2+F[E_1,E^2]^2\right).
\end{equation}
The Gauss constraints are trivialized by this second phase space reduction if we assume that the global conditions are such that axial gauge is a complete gauge fixing. The constraints $Q(\rho)\approx 0$ simplify under the second phase space reduction to the shift constraints
\begin{equation}
 C(\rho)=\int d^3x\,\rho A_3\approx 0.
\end{equation}
The Dirac bracket associated with the second phase space reduction again coincides with the Poisson bracket.

It is clear that electromagnetism and the shift symmetric dual theory are equivalent, since both are obtained as partial gauge fixings of the linking theory. This equivalence means in particular that both theories have identical observables and that their dynamics and Poisson algebras of observables coincide. The simplest way to explicitly see this equivalence is to further gauge fix the two theories to a so-called ``dictionary''. For this impose axial gauge $A_3(x)=0$ on electrodynamics and Gauss gauge $E^a_{,a}(x)=0$ on the dual theory and  assume again that the global conditions are such that this constitutes a complete gauge fixing. The condition $E^a_{,a}(x)=0$,  implies that $\phi_o$ is independent of $x^3$. Thus one solution to the Linking theory constraint $G(x)=0$, the one with the boundary condition $\phi_o(x_1,x_2,0)=0$,  is given by $\phi_o(x)=0$, so that the shift symmetric version of EM can be seen in this gauge to have the same remaining reduced constraints and variables as the dual reduction 
$A_3(x)=0$ on electrodynamics.
The two reduced phase spaces consists of field configurations $(A,E)$ that satisfy $A_3(x)=0$ and $E^a_{,a}(x)=0$. 

We thus see that the two theories have identical observables, since observables can be identified with functions on reduced phase space. We assumed global conditions, such that the Dirac matrix $M(x,y)=\delta(x,y)_{,3}$ is invertible, so the two Dirac brackets that are obtained through the two phase space reductions coincide (this Dirac bracket does however not coincide with the Poisson bracket anymore, since we are in reduced phase space). This implies that the observable algebras of the two theories coincide. The equivalence of dynamics follows from $H|_{G\equiv 0}=H_{dual}|_{G\equiv 0}$, which follows from solving the Gauss-constraint for $E^3$ and plugging this solution back into the electrodynamics Hamiltonian. 

The equivalence between electromagnetism and the shift symmetric dual theory allows us to describe the same physics (the observable, their algebra and their time evolution of free electromagnetic waves)  with two different gauge theories. The following table compares where the computational difficulties of these two descriptions lie:
\medskip

\begin{tabular}{|p{2.8cm}||p{5.8cm}|p{5.8cm}|}
 \hline
 &Maxwell description&Shift symmetric description\\
 \hline\hline
 initial value, constraint surface &$E^a_{,a}(x)=0$ is differential, the solution: $E^3(x)=F[E^1,E^2;x)$, is non--local  & $A_3(x)=0$ is algebraic, the solution $A_3(x)=0$ is local\\
 \hline
 gauge invariance condition & $\{O[A,E],E^a_a(x)\}\equiv 0$ is a differential condition & $O[A,E]$ independent of $E_3$ is a trivial condition \\
 \hline
 Hamiltonian & $H$ local and quadratic in $(A,E)$ & non--local and quadratic in $(A,E)$, but local equation for determining $H_{dual}$ \\
 \hline
\end{tabular}
\medskip

This example shows how duality of gauge theories can be used to trade a simple Hamiltonian, but complicated description of observables, on the Maxwell side for a complicated Hamiltonian (the Hamiltonian is the solution to an equation), but extremely simple description of observables, on the shift symmetric side. The analogue of this is one of the main motivations for Shape Dynamics: the complicated non--linear and differential Hamilton constraints of the ADM description of gravity is traded for a very simple algebraic set of constraints at the price of a complicated Hamiltonian, which is again given as the solution of a defining equation. This table omits the description through the ``dictionary theory,'' i.e. Maxwell theory in axial gauge. This theory has a nonlocal initial value problem and a nonlocal Hamiltonian and unlike the gauge descriptions also posses a complicated Dirac bracket.

A curious observation is that the Hamiltonian of the shift symmetric description is still quadratic. This means that not only the Maxwell description, but also the shift symmetric description can be quantized using the usual techniques for quadratic field theories. 

\subsection{Boundary/bulk and bulk/bulk dualitites}

Let us now return to the global degrees of freedom that we have ignored so far. In particular, we consider a boundary at the surface $x_3=0$, because these boundary conditions illustrate aspects of the relationship between the GR/SD duality and the AdS/CFT duality. The bulk variation of the Gauss constraints and the Maxwell Hamiltonian does not have any boundary dependence if one adds the following boundary terms:
\begin{equation}
 G_B(\Lambda) = -\int d^2x n_a \Lambda E^a,\,\,\,\,H_B=-\frac 1 2 \int d^2xn_ a \epsilon^{abc} A_b B_c.
\end{equation}
The variation of the ``regulated'' constraints $\bar G(\Lambda)=G(\Lambda)+G_B(\Lambda)$ and ``regulated'' Hamiltonian $\bar H=H+H_B$ have no boundary dependence, so the boundary terms $G_B(\Lambda)$ can be interpreted as conserved charges. Electromagnetism is not holographic, i.e. we can not describe the electromagnetic waves in the bulk completely in terms of the charges $G_B(\Lambda)$; rather for each value of the $G_B(\Lambda)$ there is an infinite number of gauge--{\bf{in}}equivalent solutions to Maxwell's equations in the bulk that produce these boundary charges. However, each of these bulk solutions  allows us to relate the boundary charges to integrals over surfaces $x^3=x^3(x^1,x^2)$ in the bulk by integrating the Gauss constraint:
\begin{equation}
 \begin{array}{rcl}
   G_B(\Lambda)&=&\int_{x^3=x^3(x^1,x^2)} d^2x_{12}\,\Lambda(x_1,x^2)E^3(x^1,x^2,x^3(x^1,x^2))\\
  &&+\int d^2x_{12}\int_0^{x^3(x_1,x^2)}ds\, \Lambda(x^1,x^2)(E^1_{,1}+E^2_{,2})(x^1,x^2,s).
 \end{array}
\end{equation}
A similar construction can be performed for the shift symmetric theory. The dual Hamiltonian again acquires the boundary term $H_B=-\frac 1 2 \int d^2x_a \epsilon_{abc} A_b B^c$, but independence of the boundary values of $E^a$ also requires that the integration constants in $F[E^1,E^2;x)$ do not depend on the boundary values. Then one can consider Lagrange multipliers $\rho$ with distributional dependence on the boundary, which yields the charges $C_B(\sigma)=\int d^2x \rho A_3$, where $\rho$ is now a density of weight one on the boundary. Any solution to the shift symmetric theory automatically satisfies $A_3(x)=0$, so the boundary charges $C_B(\rho)$ can be trivially related to bulk charges
\begin{equation}
 C_B(\rho)=\int d^2x_{12}\,\rho A_3(x^1,x^2,x^3(x^1,x^2)).
\end{equation}
This is a first example that exhibits the significant differences of usual bulk/boundary dualities and dynamical bulk/bulk equivalence. Another difference is that dynamical bulk/bulk equivalence is not one-one, but one-many. All that is required is a gauge-fixing surface that is itself first class. E.g. instead of axial gauge, we could have imposed Coulomb gauge $A_{a,a}\equiv0$. The dual theory for this gauge can be constructed along the same lines as before, but using the generating functional
\begin{equation}
 F=\int d^3x \left(\tilde A_a(E^a+\phi_{,a})-\tilde \pi \phi\right)
\end{equation}
for the best matching canonical transformation. 

\section{Short Description of Shape Dynamics}\label{sec:SD}

The simplest way to explain Shape Dynamics is to stat with the ADM description of gravity \cite{ADM} and trade refoliation symmetry for local spatial Weyl symmetry using the linking theory mechanism and subsequently deparametrizing the global scale dynamics. This method of constructing Shape Dynamics is fully explained in \cite{SD:LT}. For a more thorough explanation of the mechanisms involved, see the theses \cite{SD:Sean_thesis, SD:Gomes_thesis}.  

Spatial conformal transformations are used to solve all scalar constraints of ADM, which leaves us with an ``initial data" theory , i.e. a theory without any dynamics. However, conformal transformations that are restricted to keep the total spatial volume fixed (we work for simplicity with a compact spatial manifold $\Sigma$ without boundary, so that total volume is finite) can not solve all scalar constraints, but fail to solve one global scalar constraint. This one (not yet gauge fixed) component of the scalar constraint would have been solved by adjusting the total volume. By restricting ourselves to conformal transfromations that preserve the total volume, we retain one of the scalar constraints of ADM as a generator of true dynamics. This is why proceed with best matching ADM w.r.t. spatial conformal transformations that preserve the total volume following appendix \ref{sec:BM}. For this we adjoin the conformal factor $\phi(x)$ and its canonically conjugate momentum density $\pi_\phi(x)$ to the ADM 
phase 
space and adjoin the constraints $\pi_\phi(x)\approx 0$ to the ADM constraints.

 We use the best matching generating functional
\begin{equation}\label{equ:VPCTKretschmann}
 F=\int_\Sigma d^3x\left(\Pi^{ab}e^{4\hat\phi}g_{ab}+\Pi_\phi\phi\right)
\end{equation}
to Kretschmannize w.r.t. conformal transformations that do not change the total spatial volume.  See also  appendix \ref{sec:BM} for more details on ``Kretschmannization". Capital letters in equation (\ref{equ:VPCTKretschmann}) denote transformed variables and $\hat \phi$ denotes the volume preserving part of $\phi$, i.e. $\hat \phi=\phi-\frac{1}{6}\ln\langle e^{6\phi}\rangle$, where $\langle f \rangle:=\frac{1}{V(\Sigma)}\int_\Sigma d^3x \sqrt{|g|}f$.  The best-matching canonical transformation transforms the ADM constraints into
\begin{equation}\label{equ:TphiConstraints}
 \begin{array}{rcl}
  \left.TS(N)\right|_{\pi_\phi=0}&=&\int_\Sigma d^3x\,N\left(-8\Delta \hat\Omega+\left(\frac 1 6 \langle\pi\rangle^2-2\Lambda\right)\hat\Omega^5+R\hat\Omega-\frac{\sigma^a_b\sigma^b_a}{|g|}\hat\Omega^{-7}\right)\\ \\
  TH(v)&=&\int_\Sigma d^3x\left(\pi^{ab}(\mathcal L_v g)_{ab}+\pi_\phi\mathcal L_v \phi\right),
 \end{array}
\end{equation}
where $\hat\Omega=e^{\hat{\phi}}$ and where we used the first class constraints $\pi_\phi(x)\approx 4(\pi(x)-\mean{\pi}\sqrt {|g|(x)})$. These arise as as the canonical transformation of the constraints $\pi_\phi\approx 0$ and can be written in the form of Kretschmannization constraints
\begin{equation}
 Q(\rho)=\int_\Sigma d^3x\,\rho\left(\pi_\phi-4(\pi-\langle\pi\rangle\sqrt{|g|})\right).
\end{equation}
It is straightforward to recover the ADM formulation using the gauge-fixing $\phi\equiv 0$ and performing a phase space reduction by setting $(\phi=0,\pi_\phi=4(\pi-\langle\pi\rangle\sqrt{|g|}))$.

Shape Dynamics is constructed by imposing the gauge fixing condition $\pi_\phi\equiv 0$ and performing the corresponding phase space reduction. This gauge-fixing condition commutes weakly with the diffeomorphism constraints and the Kretschmannization constraints, i.e. these constraints are not gauge fixed. However, the condition $\pi_\phi\equiv 0$ simplifies the diffeomorphism  and Kretschmannization constraints to
\begin{equation}\label{equ:linearConstraints}
 \begin{array}{rcl}
   H(v)&=&\int_\Sigma d^3x\,\pi^{ab}(\mathcal L_v g)_{ab}\\
   D(\rho)&=&\int_\Sigma d^3x\,\rho\left(\pi-\langle\pi\rangle\sqrt{|g|}\right).
 \end{array}
\end{equation}
These constraints generate spatial diffeomorphisms and spatial conformal transformations that leave the total spatial volume invariant.

The gauge-fixing is complicated by the fact that $\hat \phi$ appears instead of $\phi$ in the first line of equation (\ref{equ:TphiConstraints}). This implies that the gauge fixing condition $\pi_\phi\equiv 0$ is reducible, because
$$
  M=\int_\Sigma \pi_\phi
$$
Poisson commutes with $\hat \phi$. The construction of Shape Dynamics is complicated by the appearance of $\hat \phi$. Let us first assume $\hat \phi = \phi$, then the first line of equation (\ref{equ:TphiConstraints}) simplifies to the usual Lichnerowicz-York equation:
\begin{equation}
 8\Delta \Omega=\left(\frac 1 6 \langle\pi\rangle^2-2\Lambda\right)\Omega^5+R\Omega-\frac{\sigma^a_b\sigma^b_a}{|g|}\Omega^{-7}.
\end{equation}
This equation has a unique positive solution $\Omega_o[g,\pi;x)$ for almost all initial data, which implies a unique $\phi_o[g,\pi;x)$. This implies that whenever $\Omega_o[g,\pi;x)=\hat\Omega_o[g,\pi;x)$ we can express the constraint surface equivalently as $e^{6\hat \phi(x)}\approx e^{6\hat\phi_o[g,\pi;x)}$. However, if the solution $\Omega_o[g,\pi;x)\ne\hat\Omega_o[g,\pi;x)$, i.e. the solution to the LY equation is not total volume-preserving, then there is no solution to the first line of equation (\ref{equ:TphiConstraints}), which is nothing but the statement that there is a remaining constraint that can be expressed as a volume constraint
\begin{equation}\label{equ:volumeConstraint}
 \int_\Sigma\sqrt{|g|}\left(1-e^{6\phi_o[g,\pi]}\right)\approx 0.
\end{equation}
This expression is independent of $\phi$ and thus not gauge-fixed by $\pi_\phi\equiv 0$. We thus see that imposing the constraints $\pi_\phi\equiv 0$ leads to the phase space reduction $(\hat\phi=\hat\phi_o[g,\pi;x),\pi_\phi(x)=0)$. This trivializes all scalar constraints of ADM except for the volume constraint (\ref{equ:volumeConstraint}), which remains unaffected by the phase space reduction.

We have thus constructed a theory with spatial diffeomorphism and spatial conformal constraints that preserve the total spatial volume (equation \ref{equ:linearConstraints}) and a constraint  on the total volume (equation \ref{equ:volumeConstraint}). This is not yet a true theory of Shape Dynamics, because (1) the total volume is still observable and (2) there is no dynamics, just a Hamilton constraint. These two problems can be overcome by realizing that the volume constraint is of the form of a time reparametrization constraint $p_t-H(t)\approx 0$, if one identifies the total volume with the momentum conjugate to $t=\tau$, where we identify time as York-time $\tau=\frac 3 2 \langle \pi\rangle$. This choice of time eliminates the only two physical degrees of freedom that still have scale: $\langle\pi\rangle,V$. This allows us to go to a completely scale--invariant theory by deparametrizing the system w.r.t. $\tau$, which also eliminates its conjugate momentum $V$. We thus obtain the constraints
\begin{equation}\label{equ:Constraints17}
 \begin{array}{rcl}
  D(\rho)&=&\int_\Sigma d^3x \,\rho\left(\pi-\frac 2 3 \tau\sqrt{|g|}\right)\\ \\
  H(v)&=&\int_\Sigma d^3x \pi^{ab}(\mathcal L_v g)_{ab},
 \end{array}
\end{equation}
which generate unrestricted spatial diffeomorphisms and conformal transformations. The evolution in $\tau$ is generated by the physical Hamiltonian
\begin{equation}\label{equ:Hamiltonian18}
 H_{SD}(\tau)=\int_\Sigma d^3x \sqrt{|g|} e^{6\phi_o[g,\pi,\tau]}.
\end{equation}
The constraints \ref{equ:Constraints17} generate unrestricted conformal transformations\footnote{$\{D(\rho),.\}$ generates unrestricted conformal transformations of the metric as gauge transformations while leaving the trace--free part of the metric momenta invariant. The constraint $D(\rho)=0$ on the other hand completely determines the trace part of the metric momenta.} and the Hamiltonian \ref{equ:Hamiltonian18} is invariant under unrestricted conformal transformations, but it depends on an external time variable. It is thus a dynamical theory of spatial conformal geometry with time--dependent Hamiltonian. The evolution of the equivalence class of conformal geometries in this description is equivalent with ADM in CMC gauge \emph{if one requires the time variable to be York time and the ADM volume to be the  volume given in \ref{equ:Hamiltonian18}.}

Let us conclude the description of SD with the remark that matter can be easily included into the system. For standard matter, one simply extends the best matching canonical transformation by a transformation that leaves all matter degrees of freedom (and their canonically conjugate momentum densities) unchanged (for details see \cite{SD:matter}).

\section{Questions and Answers}\label{sec:QA}

In this section, we list a representative selection of frequently asked questions about Shape Dynamics. We will first give a short summarizing answer, and in some cases follow the shorter answer by a more detailed discussion of the given question. 

\begin{enumerate}

 \item {\it What is the precise relation between SD and ADM?}

{\bf Short answer:}

 Both theories exist on the \emph{same phase space}, i.e. the ADM phase space with the canonical Poisson bracket thereon. Both theories posses the same observables and observable Poisson algebra and both theories have spatial diffeomorphisms as a gauge symmetry. The difference is that the relativity of simultaneity of GR is replaced with an absolute notion of simultaneity and relativity of spatial scale. This is why SD can {\emph not} be obtained as any gauge-fixing of ADM.

{\bf Long answer:}

Both Shape Dynamics and ADM are obtained from a larger Linking theory, possessing a conformal field degree of freedom in addition to the usual metric degrees of freedom. Each of two distinct gauge fixings of this Linking Theory produce each of the dual theories: ADM and SD. In the SD side one solves the extended scalar constraint and is left with a very simple set of local first class constraints, all linear in the momenta, which enables one to implement their action very simply as vector fields in configuration space. Furthermore the constraint algebra forms a true algebra, as opposed to ADM. In a concrete sense, the local constraints all acquire the simple characteristics usually attributed solely to the 3-diffeomorphism constraint in ADM.  The difficulty in the SD side arises upon solving the extended scalar constraint: one obtains a non-local evolution Hamiltonian. The upshot is that we have simple local constraints and the spatial Weyl transformations acting as a gauge group.

\item{\it Are there solutions of ADM that are not solutions in SD? And vice-versa? }

{\bf Short answer:} 
The answer requires distinction: Locally ADM and SD can not be distinguished. However it is known that there exist solutions to GR that can not be translated into SD due to global obstructions. The converse is also true: there are solutions of SD that cannot be translated into solutions of GR because the reconstructed line element does not form a non-degenerate space-time.

{\bf Long answer:}

Let us first of all, set notation and  denote by gauge$_{\mbox{\tiny ADM}}$ and gauge$_{\mbox{\tiny SD}}$ the symmetry groups of each theory. The first are the phase space equivalent of space-time diffeomorphisms, the latter are foliation and volume preserving Weyl transformations and foliation preserving diffeomorphisms. 

One does not need to impose CMC in order to solve the LY equation (see below), and that fact allows us to define the SD Hamiltonian everywhere in phase space.\footnote{Uniqueness is broken for configurations where $\pi^{ij}=0$ on closed manifolds. } However, constancy of the trace of the momentum \emph{is} a constraint in SD, which needs to hold for \emph{any solution}.  Some $(g,\pi)$ that satisfies  the scalar constraint will automatically have the trivial LY conformal factor $\phi_o[g,\pi](x)=0$.  If this, along with the 
diffeomorphism constraint and the CMC constraint are  maintained for all instants, then such field configurations are solutions for SD and ADM.

Let us define a \emph{weakly CMC foliable space-time} if the space-time has a \emph{global} constant mean curvature foliation for which the lapse is not identically zero for any time $t$. The usual notion of CMC-foliable - which we will call \emph{strongly CMC foliable} - requires the  lapse to never reach zero at any single point in space. Strong CMC foliability is required for an ADM solution to form a space-time (since we can invert the 3-metric everywhere to obtain the 4-metric), and also for this space-time to have a Shape Dynamics dual. The existence of solutions of GR that are not  weakly CMC foliable  signals that these solutions will \emph{not} be solutions for Shape Dynamics. That is, this ADM solution is not gauge$_{\mbox{\tiny ADM}}$-equivalent to a solution of SD.

 However, one can still have a Shape Dynamics solution, which, once we use the solution of the lapse-fixing equation to translate back to a space-time, does not correspond globally to a space-time solution of the Einstein equations. For instance, suppose that the reconstructed space-time possesses only a weak CMC foliation. This means that there is no gauge$_{\mbox{\tiny ADM}}$-equivalent space-time that does not have a degenerate metric. In fact, a Shape Dynamics exact solution to the spherically symmetric case contains an event horizon but no black hole interior (and no event horizon) \cite{SD_Birkhoff} and exhibits exactly the behavior described in this paragraph. 

The picture that emerges for Shape Dynamics is that of a continuous curve in phase space. It is this continuity that matters, and not space-time continuity of the rebuilt line element. In fact, no singularities arising from the focusing theorem can appear in this picture.  That is because the expansion scalar is not a quantity physically meaningful in the reduced conformal phase space. This is confirmed by the well-known singularity avoindance properties of maximal slicings.

 \item {\it What is the role of CMC gauge in SD?}

{\bf Short answer:} 

Shape Dynamics does not allow for the concept of foliations, and in fact permits the conjugate momenta to have non-constant trace off-shell. However, to interpret a Shape Dynamics solution as an ADM solution, one needs to go to a gauge in which the trace of the extrinsic curvature is a spatial constant. Conversely, to interpret a solution of the ADM system as a shape dynamics solution, one needs to impose CMC gauge, globally (on the entire space-time); this is of course only possible if the ADM solution is globally CMC foliable.\footnote{Any given space-time, around a given point there is a local patch which is CMC foliable. Hence the question becomes relevant only globally.  } 

{\bf Long answer:}

The very notion of a foliation of spacetime is foreign to Shape Dynamics: The ontology that underlies Shape Dynamics is that of a spatial conformal geometry that evolves in time. In fact, asking about a preferred foliation in Shape Dynamics is as meaningless as asking for a preferred scale in General Relativity. However, if one imposes CMC gauge in General Relativity, then the Hamilton constraints determine a particular spatial scale. Similarly, imposing the gauge $\hat \Omega(x)=0$ on Shape Dynamics allows us to use the conformal constraints to determine a CMC foliation.

Moreover, the construction of Shape Dynamics does not require CMC gauge, as we mentioned above regarding the solution of the  Lichnerowicz-York  equation.  This is because the construction of shape dynamics through the linking theory includes a step in which $\pi(x)$ is replaced by $\langle \pi \rangle\sqrt{|g|}(x)$ before constructing the SD Hamiltonian by solving the Lichnerowicz-York equation. We are thus always solving a Lichnerowicz-York equation with constant trace, for any initial data, whether it satisfies the CMC condition or not. The point is that when the CMC conditions \emph{is enforced}, the solution will coincide with that in usual ADM in CMC gauge.  The existence and uniqueness conditions for a positive solution to the Lichnerowicz-York equation with CMC data are thus directly applicable to the construction of the Shape Dynamics Hamiltonian over the entire phase space.

\item {\it What is the difference and similarity between ADM in CMC gauge and Shape Dynamics?}

{\bf Short answer:} 

  Shape Dynamics possess an unreduced local phase space with canonical Poisson brackets and the usual representation of the diffeomorphism constraint, for general functionals of the metric and metric momenta. ADM in CMC possesses a reduced non-local\footnote{In a sense to be made precise in the appendix.} phase space with non-canonical Poisson brackets. Furthermore the diffeomorphism constraint for ADM in CMC gauge requires an extraneous  background structure to generate diffeomorphims.

{\bf Long answer:} 

Both theories possess a global evolution Hamiltonian. From the ontology of Shape Dynamics, a space-time solution will only have physical significance if it possesses a conformally invariant dual. As stated in the response to  `{\bf Are there solutions of ADM that are not solutions in SD? And vice-versa?}', this will only happen if a space-time has a global strong-CMC foliation, which is when it matches Shape Dynamics. Thus ADM in CMC can be seen, on its own rights, as the reduced theory for Shape Dynamics, provided one does not demand of a solution curve of ADM in CMC to rebuild a space-time line element.
  
However,  Shape Dynamics' phase space is unreduced, being parametrized by $(g_{ij},\pi^{ij})$ with the usual Poisson bracket. Explicit spatial coordinate independence is contained in the usual form of the diffeomorphism constraint. It is a spatial-Weyl gauge theory. ADM in CMC gauge  on the other hand, requires a reduced phase space, whose parametrization we denote in the appendix by $(\rho_{ij},\sigma^{ij})$. Presented with a general functional of the metric and metric momenta $F[g,\pi]$ , one must first project down to the reduced phase space $\tilde F[\sigma, \rho]$, and in that sense the theory is non-local.  The Poisson bracket between $\sigma_{ij}$ and $\rho^{ij}$ is not canonical, but possesses a traceless projection operator.  Furthermore the momentum constraint generates diffeomorphisms  \emph{ on already conformally invariant functionals}    $F[\sigma, \rho]$ of the variables $\sigma^{ab}, \rho_{ab}$(and even then, only those whose flux is divergenceless). For a general functional of the metric 
and metric momenta, $F[g,\pi]$, one requires auxiliary structure to project down to  $ F[\sigma, \rho]$. The replacement $(g_{ij}\pi^{ij})\mapsto (\rho_{ij}, \sigma^{ij})$ is coordinate-dependent (due to the different density weights). Thus one needs a reference density weight to unambiguously define the projection.   One of the main advantages of Shape Dynamics is that one has diffeomorphism invariance intact for the full variables $g_{ab}, \pi^{ab}$. 

These are all classical differences between the two theories. Upon quantization, Shape Dynamics has a richer topological structure since one does not require phase space reduction and no Dirac brackets. The fact that the Poisson bracket of the variables are canonical also simplifies canonical quantization. 

 \item {\it How does the initial value problem for Shape Dynamics compare to  the  initial value problem of ADM?}

{\bf Short answer:}

The initial value problem for Shape Dynamics consists solely of finding transverse-traceless momenta. For ADM, the only general way of solving the initial value problem consists of using the conformal York method, which takes as input transverse-traceless momenta, and finds suitable metrics by adjusting the conformal factor to be given by the solution of the LY equation.  

{\bf Long answer:}

In ADM, the initial data problem is most generally attacked by the York conformal method. This requires transverse-traceless momenta (which already implies the solution of a second order differential equation) and a  conformal decomposition of the metric. This data is then  input into the scalar constraint, seen now as an equation for the conformal factor in terms of the TT momenta and the conformal class of the metric. Having obtained such a solution for the conformal factor, one then reintroduces this into a representative of the conformal class of the metric, thereby obtaining initial data $(g^o_{ij},\pi^{ij}_{\mbox\tiny TT})$.
 
In Shape Dynamics, the local constraints are given by the usual diffeomorphism constraint and the CMC constraint. The leftover global constraint can be posed either as reparametrization constraint or as \emph{an evolution Hamiltonian}.

 Let us  stress that unlike either the York method or ADM in CMC,  in Shape Dynamics one need not go to reduced phase space  to obtain the dynamical evolution of any functional of the canonical variables.  One can impose the constraints only weakly, which means that one can calculate Poisson brackets and impose the constraints only after variation. As stressed before, the  evolution Hamiltonian can be obtained from the extended scalar constraint $t_\phi S(x)=0$,  which does \emph{not} require transverse-traceless momenta or metrics to be in the York conformal section. This represents an  extension of the usual York method and  LY equation for the conformal factor.  

We should add that the simplicity of the initial data, and this crucial modification of the  LY equation (which in our case does not require TT momenta, which is why we prefer to refer to it as ``the extended Hamiltonian") could be instrumental in exploring new exact solutions of gravity.

\item {\it How does the  the relative status of simultaneity in Special Relativity enter the description of Shape Dynamics? }

{\bf Short answer:} Special relativity enters the picture in Shape Dynamics because for solutions of Shape Dynamics over flat space, with the usual asymptotically flat boundary conditions, there remains a redundancy of generators of time displacement. These turn out to exactly match the generators of boosts.
In this way, Special Relativity emerges for Shape Dynamics in the same way it emerges from the ADM description of GR: Given asymptotically flat boundary conditions, both descriptions posses a time--independent solution given by a flat spatial metric and vanishing metric momenta. This solution has a 10--parameter group of global symmetries, which is the Poincar\'e group of Special Relativity. 

This is the sense in which Lorentz invariance enters the picture in a global manner, as in the dynamics of the (flat) metric. In a local manner, one can always rebuild a local patch of space-time that will have Lorentz invariance for local observers. 

\item {\it How can approximately FLRW cosmology be equivalent to a theory in which scale has no meaning? }

{\bf Short answer:} The resolution of this puzzle is presented in \cite{Barbour:2013goa}. The argument is, loosely speaking, that there is no objective dynamics in pure gravity FLRW, say on $S^3$, since the only rod available in this model is the total volume. Compatibility with observation is achieved by including a sufficient number of degrees of freedom to observe expansion: E.g. a photon gas (modeling the CMB) and an additional oscillator (modeling the reference scale on a spectrometer). The objective predictions derived from this system coincide with the objective predictions of standard FLRW. 

{\bf Long answer:} The argument made in \cite{Barbour:2013goa} is based on the fundamental physical assumption that the description of a closed classical universe must not require any absolute dimensionful scale. All scales must be defined within the universe itself. In fact, international units are precisely defined as such ratios within the universe.\footnote{There is of course the caveat that the modern definition of units uses quantum reference systems, while the argument in \cite{Barbour:2013goa} is laid out for a closed classical system.} Thus, if, as in pure FLRW, the only degree of freedom is the total spatial volume, there is no objectively definable evolution; the system consists of a single degree of freedom (and its momentum) that does not evolve w.r.t. any physical clock. To be able to define such a clock within the universe, one needs more degrees of freedom in the system, such as e.g. matter or shape degrees of freedom (e.g. extending pure FLRW on $S^3$ to pure Bianchi IX). However, the above construction shows that the objective dynamics, i.e. the evolution of objectively definable ratios within the model universe, of the completely scale--invariant theory coincides with the objectively definable ratios within the description with restricted scale--invariance (i.e. the theory with volume). This coincidence of the {\bf objective} physical predictions of the two descriptions reconciles full scale--invariance with the apparent expansion of the universe as it is deduced from redshift. The derivation of redshift within Shape Dynamics is the content of a soon to appear note in which the York time dependence of the ratio $\frac{\omega_{light}(k_o)}{\omega_{oscillator}(k_o)}$, where $\omega_{light}(k_o)$ denotes the dispersion relation of light and $\omega_{oscillator}(k_o)$ denotes the dispersion relation of a massive scalar field, is derived in the Shape Dynamics description of gravity. Thus, light which  is emitted with the same frequency as a material reference oscillator at some initial value of York time will appear to have a shifted frequency when compared with the same material reference oscillator at later York time. This dependence shows how the observed redshift is compatible with the Shape Dynamcis description of gravity.

\item {\it How can a nonlocal Hamiltonian describe local propagation ?}

{\bf Short answer:}

A sufficient condition for local evolution is that there is an equivalent formulation of the theory that is manifestly local, i.e. the Hamiltonian and constraints are integrals over local densities. If one would not include this equivalence in the definition of locality, one would not be able to call any theory local, since it is always possible to describe a equivalently with an apparently nonlocal set of constraints or nonlocal Hamiltonian. The SD evolution is local in this sense, since SD is manifestly equivalent to GR evolution whenever a CMC foliation is available.

{\bf Long answer:}

In many cases, however not in SD, there is a simpler answer to the question that does not require to resort to an equivalent local description. The observation is that a sufficient condition for local evolution equations is that there is a local set of gauge fixing conditions, such that the restriction of the constraints and Hamiltonian to this gauge fixing surface coincides with manifestly local functions. This is the case in the shift-symmetric example above. The nonlocal Hamiltonian of the shift symmetric dual of the electrodynamics Hamiltonian coincides with the electrodynamics Hamiltonian on the surface where the Gauss constraints hold, which is a set of local gauge-fixing conditions for the shift symmetric theory. This simple argument does not hold in SD, because the gauge-fixing in which SD coincides with ADM in CMC gauge is nonlocal. This nonlocality is due to the fact that one does not use all scalar constraints of ADM, but only those that are linearly independent of the CMC Hamiltonian. This is why 
one has to resort e.g. to the local formulation of the linking theory to see manifest locality in SD. This local formulation is obtained by replacing $\hat \phi$ with an unrestricted $\phi$ that is accompanied by the condition $\int d^3x \sqrt{|g|}(1-e^{6\phi})=0$.

\item {\it Are there procedures for dealing with this non-locality?}

{\bf Short answer:}

Yes.

{\bf Long answer:} 

Given special CMC initial data, i.e. CMC data for which the spatial Ricci scalar is homogeneous, one can solve the Lichnerowicz--York for pure gravity on a compact manifold algebraically. This is due to the fact that all coefficients of the Lichnerowicz--York equation are homogeneous, so the maximum principle implies that the solution to the Lichnerowicz--York equation is homogeneous, which means that the solution can be found by finding the positive root of a third order polynomial. This exact solution can be used as the starting point for a systematic expansion of the Shape Dynamics Hamiltonian in spatial derivatives \cite{Koslowski:2013gua}. This provides an algebraic recursion relation for the York factor order by order.

 The situation simplifies dramatically when the spatial Ricci scalar of a CMC slice vanishes, where the exact solution to the Lichnerowicz--York equation is the positive root of a quadratic equation. The first two orders of the Hamiltonian in this case are
\begin{equation}
 H=\int d^3x \sqrt{|g|}\left(\sqrt{\frac{M}{T}}+\frac{1}{2T}\,\left(\frac{M}{T}\right)^{\frac{1}{12}}(8\Delta-R)\left(\frac{M}{T}\right)^{\frac{1}{12}}\right)+\mathcal O(\nabla^4),
\end{equation}
where $M=\frac{\sigma^a_b\sigma^b_a}{|g|}$ and $T=\frac 3 8 \tau^2-2\Lambda$. The combination of Shape Dynamics with locality suggests in this way a modified classical gravity theory: the modified Hamiltonian would be obtained by truncating the derivative expansion at a finite order and conformalizing it.

 \item {\it Why do you sometimes say volume-preserving Weyl transformations, and at other times you seem to leave it unrestricted ?}

{\bf Short answer:}

This is because Weyl transformations appear at two different places of the construction. First they appear in the construction of the linking theory, where they are restricted to preserve the total spatial volume, if the Cauchy surface is compact and second they appear as the gauge group of deparametrized SD, which is unrestricted Weyl transformations.

{\bf Long answer:} 

The reason for using spatial Weyl-transformations that preserve the total volume in the construction of the linking theory is because we want to retain one of the ADM scalar constraints as a generator of dynamics. The volume preservation condition is necessary, because conformal transformations can be used to weakly solve all ADM scalar constraints, so to retain a nontrivial generator for time evolution requires us to impose restrictions on the conformal transformations. The simplest such restriction is preservation of the total spatial volume in the compact case. It turns out that this restriction on the conformal transformations has the very nice property that best matching w.r.t. volume preserving conformal transformations yields York-scaling. This is important, because York-scaling ensures the existence and uniqueness of a generator of dynamics after symmetry trading for all interesting sets of initial data.

Using best matching w.r.t. volume preserving conformal transformations and straightforwardly applying the symmetry  trading procedure yields a theory whose constraints are volume preserving conformal transformations and a generator of constraint dynamics that can be written in the form of a volume constraint. This system is not the purest form of SD, because a global volume scale should not appear in SD. This can be remedied by identifying a parametrized dynamical system: replace the mean extrinsic curvature by time (in particular York time) and the total volume can be identified as its canonically conjugate momentum. The removal of the total volume as a physical degree of freedom yields a new system with \emph{time-dependent, unrestricted conformal constraints}, i.e. unrestricted Weyl invariance, and a time-dependent Hamiltonian. The translation of a solution of the fully conformal description into the ADM 
picture however requires one to interpret the time variable as York--time and to calculate the on--shell value of the total spatial ADM volume using \eqref{equ:Hamiltonian18}. 

 \item {\it What is the relation of the GR/SD duality with the AdS/CFT duality?}

{\bf Short answer:}

GR/SD duality is a bulk/bulk duality in the sense that it holds on any CMC Cauchy surface of ADM. A particular bulk/boundary duality can be obtained in a large CMC volume expansion, which corresponds to a boundary in time. The leading orders in this expansion coincide with a particular approach to holographic renormalization.

{\bf Long answer:}
Shape Dynamics rests on a duality between ADM and a foliation preserving  spatial Weyl gauge theory on each CMC Cauchy surface of ADM.  This is a bulk/bulk duality. However, a large CMC volume expansion of the Hamiltonian yields a very simple generator for the evolution of shape degrees, which is only valid at asymptotically very large volume \cite{SD:HJ}. It turns out that in this domain we may obtain different conformal field theories (for different types of matter). One can do a Hamilton-Jacobi expansion and explicitly check the standard results for holographic renormalization \cite{Skenderis}. In \cite{Skenderis}, this is seen as a check on the AdS/CFT duality, but SD offers a different explanation: the underlying reason is the duality with a Weyl gauge theory. Holographic renormalization recuperates the duality asymptotically because in that domain the temporal gauge used becomes the CMC gauge, for which  dynamical equivalence with SD is explicit.  

A different point of contact is the treatment of boundary charges in SD, in particular in asymptotically AdS space, which are now considered important evidence for AdS/CFT. We saw in the electromagnetism toy model that some boundary charges can be related to the dual theory. The analogous construction in SD is currently performed and seems to provide suggestive relations \cite{Lee_SD}.

\item{\it What is the relationship between Shape Dynamics and the so called Doubly General Relativity?}

{\bf Short answer:} The main result of doubly general relativity \cite{Gomes:2012hh} regards a BRST treatment of a CMC gauge fixing of ADM. The result is that in the case of pure constraints theories, such as ADM, gauge fixing terms that are also symmetry generators possess a special role in the classical BRST formulation of the gauge-fixed theory: the gauge-fixed Hamiltonian in this instance has the BRST symmetries related to both the original symmetry and to that of the gauge-fixing term. It is mathematically related to new Shape Dynamics, but a structurally  different theory, since one of the main characteristics of Shape Dynamics is that it is not gauge-fixed.

 \item {\it What is the role of Kretschmannization and best matching in SD?}

{\bf Short answer:}

Kretschmannization and Best Matching are tools for the construction of linking theories, such as the one that proves the equivalence between GR and SD.  No method of constructing Shape Dynamics without using the Kretschmannization (and the Linking Theory) is known my us. This does not mean it plays a fundamental role, but it certainly plays a technical one (so far). Linking theories and the specific construction tools are a priori not necessary to prove dynamical equivalence, but turn out to be practically very important.

{\bf Long answer:}

Kretschmannization is often called Stueckelberg mechanism and is a rather trivial process whereby additional degrees of freedom are introduced into a theory, without real dynamical consequence. However, when this process is coupled to the gauge-fixing implied by best-matching, it yields non-trivial dualities between dynamical theories. Kretschmannization by itself  extends the degrees of freedom of the theory together with additional constraints. This is what enables us to arrive at the Linking theory. Best-matching by itself implies a separation of degrees of freedom into gauge and non-gauge, and a subsequent specific form of gauge fixing. It yields  a gauge-fixed version of a theory, as in for example in Barbour's route  from ADM to ADM in CMC. When the two methods are concatenated the trading of the extended scalar constraint with the spatial-Weyl constraint ensues, yielding Shape Dynamics.

 \item {\it How does your Shape Dynamics compare to Barbour's previous formulation?}

{\bf Short answer:} 

Barbour's previous formulation differs from Shape Dynamics in many ways. It can be seen an action-based way of obtaining the conformal York equations for the initial value problem. But it has two flaws: i) it is solely a formulation of the initial value problem, and does not possess a formulation of dynamics, as other, older, but more dynamically complete formulations of ADM in CMC do (\cite{Fischer-Moncrief} and see the review in \cite{3+1_book}). It inspired Shape Dynamics in that it was derived form  first principles that are not based on space time, but on  spatial shape degrees of freedom \cite{Barbour:bm_review, Barbour:CS+V}, a point of view which  references \cite{3+1_book, Fischer-Moncrief} \emph{do not} brandish. 
 Shape Dynamics on the other hand is a \emph{gauge theory with local spatial conformal invariance} and a single distinguished Hamiltonian.

 \item {\it How does your Shape Dynamics compare to Witten's and Moncrief's quantum GR in 2+1 dimensions?}

{\bf Short answer:} 

Witten's quantization of the Chern-Simons formulation gravity \cite{Witten:1988hc} is a quantization of the initial value problem only and does not retain a generator of dynamics, which we view as essential. Moncrief's quantization is a reduced phase space quantization \cite{Moncrief:1989dx} of ADM in CMC gauge and can thus be viewed as a reduced phase space quantization of SD. In \cite{Carlip:1990kp} Carlip constructed an operator map between the two quantizations and used it to find Moncrief's Hamiltonian on Witten's Hilbert space thus completing Witten's kinematic quantization. The linearity of all local constraints in Shape Dynamics however raises the hope that it enable an unreduced quantization. 

{\bf Long answer:}

It is not difficult to use the symmetry trading mechanism to trade all scalar constraints for full conformal constraints, but the price for this is that no generator of time evolution remains. This is at least at the classical level problematic, because one ends up with a frozen theory. However, if the scalar constraints of ADM are viewed as generators of gauge transformations, then reduced phase space quantization suggests to just quantize the initial value problem. This is essentially what Witten did when quantizing 2+1. One can follow the same logic in SD and finds timeless wave-functions on Teichm\"uller space.

Moncrief's reduced phase space quantization of ADM in CMC gauge retains the York Hamiltonian and the procedure can be re-interpreted as a reduced phase space quantization of  SD. The difference between a quantization of SD and ADM in CMC gauge would however occur in a Dirac quantization program, where reduction is performed after quantization. The quantization of ADM in CMC gauge is not known. This is the point where SD has a formal advantage \cite{SD:Budd}: The constraints of SD are linear in metric momenta and can be integrated to geometric transformations. These gauge transformations can be formally quantized in a Schr\"odinger representation. However, two shortcomings remain: (1) it would be desirable to find a kinematic inner product that supports these transformations as unitary transformations and (2) a sensible quantization of the York Hamiltonian on the 2-torus $H=\int d^2x\sqrt{\frac{\sigma^a_b\sigma^b_a}{\tau^2-4\Lambda}}$ as an essentially self-adjoint operator on the kinematic Hilbert space 
should be 
constructed.

 \item {\it Is Shape Dynamics generally covariant?}

{\bf Short answer:} 

Yes, but not manifestly. More precisely: SD is equivalent to GR, which is generally covariant, but this general covariance appears only when the equations of motion hold. That this is compatible with Poincar\'e invariance can be shown rather trivially, by investigating the Shape Dynamics solution for Minkowski spacetime. 

{\bf Long answer:}

The SD evolution equations allow us to freely specify a shift vector $v$ and a local change of scale $\rho$ but determine the evolution of the spatial metric $g_{ab}$ in terms of initial data in terms of $v,\rho$. An SD trajectory is thus $(g_{ab}(t),\xi(t),\rho(t))$. Given any trajectory, we can perform a time-dependent spatial Weyl transformation $\phi:(g_{ab},\pi^{ab})\to(e^{4\phi}g_{ab},e^{-4\phi}\pi^{ab})$, such that $(e^{4\phi}g_{ab},e^{-4\phi}\pi^{ab})$ satisfies the ADM constraints along the trajectory. Moreover, we can solve the CMC-lapse fixing equation at each step in time to find a lapse $N(t)$. We thus have data $(e^{4\phi(t)}g_{ab}(t),\xi(t),N(t))$ for the entire SD trajectory, which defines a 4-metric $g_{\mu\nu}$ that satisfies Einstein's equations, which are manifestly generally covariant. 

A physical way to recover general covariance is by coupling it to a multiplet of matter fields. This allows one to reconstruct spacetime by observing how matter evolves on an SD solution. The reconstructed spacetime has precisely the metric $g_{\mu\nu}(e^{4\phi}g_{ab},\xi^a,N)$ whose reconstruction we just described.



\end{enumerate}

\section{Conclusions}

The purpose of this paper is to give answers to some of the most frequently asked questions regarding new Shape Dynamics. It aims above all to make clear the distinctions and advantages to previous formulations of gravity and its gauge fixings. The answers to the majority of these questions can be summarized in as follows:
\begin{enumerate}
 \item {\it Dynamical Equivalence:} The observable algebras of Shape Dynamics and ADM, i.e. the set of observables and their Poisson algebra, as well as the time evolution of these observables coincide. Nonetheless, there are solutions of GR which globally cannot be translated into Shape Dynamics, and the converse is also likely to hold. 
 \item {\it Difference with gauge-fixed GR and previous SD:} New Shape Dynamics is, unlike any previous formulation of GR, a gauge theory of spatial diffeomorphisms and Weyl transformations on unreduced ADM phase space. ADM in CMC has diffeomorphism invariance, but only with the addition of an auxiliary background metric. 
 \item {\it Locality and general covariance:} Shape Dynamics has a local evolution and is generally covariant, but these symmetries appear only on-shell.
 \item {\it Relation with AdS/CFT:} The bulk/bulk equivalence of Shape Dynamics and ADM reduces in a large volume limit to terms known from holographic renormalization. 
\end{enumerate}
We gave more detailed questions in section \ref{sec:QA}. Where appropriate, we gave a short answer followed by a more detailed explanation. 

To illustrate part of these relations without having to introduce technical baggage we have included a toy model (section \ref{sec:EM}): electrodynamics for which we perform symmetry trading to obtain a dual shift symmetric theory. The dictionary between this toy model and gravity is as follows: The Maxwell formulation can be understood as an analogue of ADM, the shift symmetric dual can be considered as an analogue of SD and electromagentism in axial gauge can be considered as the analogue of ADM in CMC gauge. This formulation opens the door to more complicated symmetry trading in ordinary gauge theories. 

The technical explanations in the appendix highlight some aspects of Shape Dynamics that have not  been properly addressed in the literature. These are
\begin{enumerate}
 \item Straightforward attempts to quantize Shape Dynamics as wave functions of the metric are very similar to the analogous attempts to quantize ADM in CMC gauge, but with the important difference that one works with an unreduced phase space. This means in particular that one can impose canonical commutation relations coming from the Poisson bracket rather than non-canonical commutation relations coming form a Dirac bracket.  
 \item The initial value problem for Shape Dynamics is significantly simpler than the initial value problem for ADM, since it is solved by finding transverse traceless momenta for a given metric. 
 \item We explicitly recovered Poincar\'e invaraince of a set of Shape Dynamics data that represents Minkowski space. In particular, the generators of boosts and time translations appear as solutions to the asymptotically flat lapse fixing equation, i.e. lapses in the set $\{1, x^a\}$. 
\end{enumerate}

Besides fulfilling its purpose of clarifying certain issues known in the community to a broader audience, we highlight  three important new results on which most of the clarifications are based: 1) The construction of a ``shift" dual to standard electrodynamics using the mechanism of symmetry trading. 2) The careful construction of ADM in CMC gauge as a dynamical theory and the status of diffeomorphism invariance in this theory. 3) The realization that Poincar\'e invariance holds for a particular solution of Shape Dynamics over Minkowski initial data.

\section*{Acknowledgements}

The authors would like to thank the referees for carefully reading the manuscript and suggesting important improvements to the paper.   HG was supported in part by the U.S.
Department of Energy under grant DE-FG02-91ER40674. TK was supported through the Natural Sciences and Engineering Research Council of Canada.

\begin{appendix}

\section{ADM formulation}\label{sec:ADM}

The ADM formulation \cite{ADM} of GR can be obtained from a Legender transform of the Einstein-Hilbert action on a globally hyperbolic spacetime $\Sigma\times \mathbb R$. This Legendre transform can be performed in a generally covariant way (cf e.g. \cite{Thiemann:2007zz}), but we will take a short-cut here starting with the ADM-decomposition of the space-time metric
\begin{equation}
 ds^2=-N^2 dt^2+g_{ab}(dx^a+\xi^a\,dt)(dx^b+\xi^b\,dt),
\end{equation}
where $N$ denotes the lapse, $\xi^a$ the shift vector and $g_{ab}$ the spatial metric. The Legendre transform, after discarding a boundary term, leads to primary constraints that constraint the momenta conjugate to $N$ and $\xi^a$ to vanish, which are solved by treating $N,\xi^a$ as Langrange multipliers for the secondary constraints
\begin{eqnarray}
  S(N)&=&\int_\Sigma d^3x \,N\,\left(\frac{\pi^{ab}(g_{ac}g_{bd}-\frac 1 2g_{ab}g_{cd})\pi^{cd}}{\sqrt{|g|}}-(R-2\Lambda)\sqrt{|g|}\right)\\
  H(\xi)&=&\int_\Sigma d^3x \,\pi^{ab}(\mathcal L_\xi g)_{ab},
\end{eqnarray}
where $\pi^{ab}$ denote the momentum densities canonically conjugate to $g_{ab}$. The total Hamiltonian is a linear combination of the constraints
\begin{equation}
  \mathbb{H}=S(N)+H(\xi). 
\end{equation}
The constraints $H(\xi)$ generate infinitesimal diffeomorphisms in the direction of $\xi^a$. The transformations generated by the scalar constraints $S(N)$ do not have such a simple off-shell interpretation, but on-shell, i.e. on a solutions to Einsteins equations, they generate refoliations. This is the reason why the Poisson algebra of the constraints is called the hyper-surface deformation algebra
\begin{eqnarray}
\{S(N_1),S(N_2)\}&=&g^{ab}H_b(N_1\nabla_a N_2-N_2\nabla_a N_1)\label{equ:ADM_alg:SS}\\
\{S(N),H^a(\xi_a)\}&=&-S(\mathcal{L}_\xi N)\\
\{H^a(\xi_a),H^b(\eta_b)\}&=& H^a([\xi,\eta]_a).\label{equ:ADM_alg:HH}
\end{eqnarray}
This algebra is however {\bf not} a property of the theory, but a property of the particular choice of constraint functions that we used to describe the theory.

\section{Conformal Spin Decomposition}\label{sec:Spin}

The technical insight that allowed York to develop his approach to solving the initial value problem of GR was that the spin decomposition of an arbitrary symmetric 2-tensor depends only on the conformal class of the spatial metric, which allowed York to decouple diffeomorphsim constraints (which constrain the longitudinal part of the metric momentum) and scalar constraints (which can be solved by a conformal transformation). To explain this insight in more detail, let us consider a spatial metric $g_{ab}$ and decompose a symmetric 2-tensor $\sigma^{ab}$ of density weight 0 uniquely into a transverse-traceless part, longitudinal part and trace part $\sigma^{ab}=\sigma^{ab}_{TT}+\sigma^{ab}_L+\sigma^{ab}_{tr}$. The trace part is defined as $\sigma^{ab}_{tr}=\frac 1 3 \sigma^{cd}g_{cd} g^{ab}$, the longitudinal part is defined as $\sigma^{ab}_L=g^{ca}v^b_{;c}+g^{cb}v^a_{;c}-\frac 2 3 g^{ab}v^c_{;c}$, where the semicolon denotes the covariant derivative w.r.t. $g_{ab}$, and the transverse traceless part 
satisfies
\begin{equation}
  g_{ab}\sigma_{TT}^{ab}=0\,\,\textrm{ and }\,\,{\sigma^{ab}_{TT}}_{;b}=0.
\end{equation}
The trace $\sigma=g_{ab}\sigma^{ab}$ is uniquely determined. It turns out that the vector $v^a$ is determined up to the addition of a conformal Killing vector of $g_{ab}$ by the transverse and traceless condition for $\sigma^{ab}_{TT}$, so $\sigma^{ab}_L$ and $\sigma^{ab}_{TT}$ are uniquely determined as well. 

The important insight is now that the components of the spin decomposition map into each other under a simultaneous conformal transformation of $g_{ab}$ and $\sigma^{ab}$
\begin{equation}
 g_{ab} \to \Omega^4 g_{ab}\,\,\textrm{ and }\,\,\sigma^{ab}\to\Omega^{-10}\sigma^{ab}.
\end{equation}
In particular, it can be shown that the summands of the spin decomposition transform as
\begin{equation}
 \sigma^{ab}_{tr} \to \Omega^{-10}\sigma^{ab}_{tr},\,\,\,\,\sigma^{ab}_{L}\to\Omega^{-10}\sigma^{ab}_L\,\,\textrm{ and }\,\,\sigma^{ab}_{TT} \to \Omega^{-10}\sigma^{ab}_{TT}.
\end{equation}

\section{Initial Value problem for ADM}\label{sec:IVP-ADM}

The most generic approach to the initial value problem of GR exists for CMC gauge, i.e. when the gauge condition $p=\frac{\pi}{\sqrt{|g|}}=const.$ is imposed to gauge-fix the scalar constraints of ADM, which is simplest, if we express the ADM constraints in terms of extrinsic curvature $K_{ab}=\frac{1}{\sqrt{|g|}}(g_{ac}g_{bd}-\frac 1 2 g_{ab}g_{cd})\pi^{ab}$. The gauge condition states that the trace part of $g_{ab}K^{ab}$ is a spatial constant. Inserting a constant trace part into the spin decomposition of $K^{ab}$ and using the transverse condition ${K^{ab}_{TT}}_{;b}=0$, we find that the momentum constraint $-2{\pi^{ab}}_{;b}=0$ is a constraint on the longitudinal part $K^{ab}_L$, which is required to vanish. This implies that the longitudinal part of $\pi^{ab}$ vanishes, so if the CMC condition and momentum constraints are satisfied, one can write the metric momenta the sum of a transverse traceless part $\pi^{ab}_{TT}$ and a spatially constant trace part
\begin{equation}
 \left.\pi^{ab}\right|_{{\pi^{ab}}_{;b}=0,p=const.}=\frac 1 3\, p\, g^{ab}\sqrt{|g|}+\pi^{ab}_{TT}.
\end{equation}
The conformal invariance of the spin decomposition now ensures that any conformal transformation of a solution to the momentum constraint is still a solution to the momentum constraint, so one can follow Lichnerowicz and York and consider a conformal transformation of the scalar constraints, which after division by $\Omega \sqrt{|q|}$ reads
\begin{equation}\label{equ:scale}
 8 \Delta_g \Omega = \left(\frac 3 8 \tau^2-2\Lambda\right)\Omega^5+R\,\Omega-\frac{\pi^{ab}_{TT}\pi^{TT}_{ab}}{|g|}\Omega^{-7},
\end{equation}
where $\tau=\frac 2 3 p$. Solutions to elliptic equations of the form $\Delta \Omega=P(\Omega)$ closely related to the positive roots of $P(\Omega)$. In particular, existence of a bounded positive solution $\Omega(x)$ follows from the existence of positive constants $0<\Omega_i<\Omega_s$ that satisfy $P(\Omega_i)<0<P(\Omega_s)$, which bound the solution as $\Omega_i<\Omega(x)<\Omega$. To investigate the positive roots of the RHS of (\ref{equ:scale}), we multiply it with $\Omega^7$, which yields a third oder polynomial in $\Phi=\Omega^4$:
\begin{equation}
  \left(\frac 3 8 \tau^2-2\Lambda\right)\Phi^3-R\,\Phi^2-\frac{\pi^{ab}_{TT}\pi^{TT}_{ab}}{|g|}=\left(\frac 3 8 \tau^2-2\Lambda\right)(\Phi-\alpha)(\Phi-\beta)(\Phi-\gamma).
\end{equation}
A detailed inspection of the roots of this polynomial reveal that for $(\frac 3 8 \tau^2-2\Lambda)>0$ one can use theorems for elliptic differential equations that guarantee existence and uniqueness of the solution to (\ref{equ:scale}) except for isolated non-generic cases. Combining the transverse-constant trace condition with the conformal transformation that solves the scalar constraints leads to the York-procedure for constructing generic initial data for General Relativity as the following recipe. 
\begin{enumerate}
 \item Choose an arbitrary trial metric $\tilde g_{ab}$ and an arbitrary trial $\tilde K^{ab}$.
 \item Derive the transverse-constant trace part of $\tilde K^{ab}_{TT}+$. This solves the momentum constraints for $\pi^{ab}$.
 \item Solve the scale equation (\ref{equ:scale}) for $\tilde g_{ab}$ and $\tilde \pi^{ab}_{TT}+\frac 1 3\, p\, g^{ab}\sqrt{|g|}$ and hence find the scale factor $\Omega$ that solves the scalar- and diffeomorphism- constraints in terms of the rescaled data.
\end{enumerate}

\subsection{Initial value problem for SD}

The initial value problem for ADM in CMC gauge coincides by construction with the initial value problem of Shape Dynamics in ADM gauge. The initial value problem for SD is however much simpler, because the local Shape Dynamics constraints are linear in the momenta and can be solved by a spin decomposition of the metric momenta. General initial data for Shape Dynamics can thus be obtained as follows:
\begin{enumerate}
 \item Choose an arbitrary spatial metric $g_{ab}$ and an arbitrary symmetric tensor density $\pi^{ab}$.
 \item Project onto the transverse traceless part of $\pi^{ab}$; this gives initial data $(g_{ab},\pi^{ab}_{TT})$.
\end{enumerate}
Notice that the momentum constraint can be solved before or after solving the conformal constraint, because the spin decomposition of $\pi^{ab}$ is $L^2$-orthogonal.

\section{ADM dynamics in CMC gauge}\label{sec:ADM-CMC}

To the extent of the authors knowledge, the construction of the ADM in CMC dynamical theory presented here is new. For related constructions, see \cite{Isham} or \cite{3+1_book}. 

To fully gauge fix the CMC constraint, it proves easier to do so from the ground up, by first introducing the following separation of variables:
\be\label{equ:new_variables}
(g_{ab},\pi^{ab})\rightarrow (|g|^{-1/3}g_{ab}, |g|,|g|^{1/3}(\pi^{ab}-\frac{1}{3}\pi g^{ab}),\frac{2\pi}{3\sqrt{ |g|}})=:(\rho_{ab},\varphi, \sigma^{ab},\tau)
\ee
where we have denoted the determinant of the metric by $|g|$, and used $\varphi$ to denote the physical conformal factor of the metric, as opposed to the auxiliary (Stuckelberg) conformal factor $\phi$. However, to simplify matters, we choose a reference metric $\gamma_{ij}$ to determine a reference density weight.\footnote{The metric $\gamma_{ij}$ can be given by a homogeneous metric depending on the topology of the space in question. E.g. if we restrict our attention to connected sums of $S^3, S^2\times S^1$ and $T^3$, then the reference metric can be obtained from the round metric for $ S^3$, the flat one for $ T^3$ and the Hopf one for $S^2\times S^1$. If we would like to be completely general, it would be better to use a reference section of the conformal bundle, for example given by metric in the conformal class of each $g_{ij}$ that has constant scalar curvature. In this case, we would have the reference metric as a functional of $g_{ij}$, i.e. $\gamma_{ij}[g]$. } Then we define a \emph{scalar} 
conformal factor $\phi:=|g|/|\gamma|$, and replace, in \eqref{equ:new_variables} $\varphi\rightarrow \phi$. Nonetheless, this still defines a physical (now scalar) conformal factor, which has the same conformal weight as the densitized version.

  The variable $\tau$ will give rise to what is commonly known as \emph{York time}. The inverse transformation from the new variables to the old  is given by:
 \be\label{equ:inv_transf}\pi^{ab}=\phi^{-1/3}(\sigma^{ab}+\frac{\tau}{3}\rho^{ab})~, ~~ \mbox{and} ~~g_{ab}=\phi^{1/3}\rho_{ab}\ee

 The non-zero Poisson brackets  are given by 
\be\begin{array}{rl}
\{\rho_{ab}(x),\sigma^{cd}(y)\}&=(\delta_{(a}^c\delta_{b)}^d-\frac{1}{3}g_{ab}g^{cd})\delta(x,y)\\
\{\phi(x),\tau(y)\}&=\delta(x,y)\\
\{\sigma^{ab}(x),\sigma^{cd}(y)\}&=\frac{1}{3}(\sigma^{ab}\rho^{cd}-\sigma^{cd}\rho^{ab})\delta(x,y)
\end{array}\ee
 The important point however is that what we will designate as ``non-physical" variables $\tau$ and $\phi$, commute with the physical modes $\sigma^{ab}$ and $\rho_{ab}$. 

We note that since the Poisson bracket between the $\sigma$ variables does not vanish, schematically $\{\sigma, \sigma\}\neq 0$, and $\{\sigma, \rho\}\neq 1$, $\sigma$ and $\rho$ do not form,   in the strictest sense of the term, a canonical pair. The Poisson bracket in terms of these variables takes the generalized schematic form:
\be \{F,G\}= \{\xi^\alpha, \xi^\beta\}\left(\delta_\alpha F\delta_\beta G- \delta_\beta F\delta_\alpha G\right)
\ee 
where we use the DeWitt generalized summation convention. In our case this becomes
\begin{multline}\label{equ:modified_PB}
\int d^3 x\Big(\diby{F}{\phi}\diby{G}{\tau}-\diby{G}{\phi}\diby{F}{\tau}+
(\delta_{(a}^c\delta_{b)}^d-\frac{1}{3}g_{ab}g^{cd})(\diby{G}{\sigma^{ab}}\diby{F}{\rho_{cd}}-\diby{F}{\sigma^{ab}}\diby{G}{\rho_{cd}})\\
+\frac{1}{3}(\sigma^{ij}\rho^{kl}-\sigma^{kl}\rho^{ij})(\diby{G}{\sigma^{ij}}\diby{F}{\sigma^{kl}}-\diby{F}{\sigma^{ij}}\diby{G}{\sigma^{kl}})\Big)
\end{multline}

In the reduced phase space, the momentum constraint also decouples from the conformal factor, as we now show, in the smeared form of the diffeomorphism constraint \footnote{It is also possible to show that the usual form of the momentum constraint $\nabla_i \pi^{ij}$ for the decomposition \eqref{equ:new_variables}, once one takes into account that $\tau$ is a spatial constant and $\sigma^{ij}$ is a traceless tensor of density weight $5/3$, yields $ \nabla_i\sigma^{ij}=\partial_i\sigma^{ij}+\hat\Gamma^j_{\phantom{i}ik}\sigma^{ik}$ where $
\hat\Gamma$ are the Christoffel symbols for the variable $\rho_{ij}$.  }:
\begin{multline}
\int d^3 x ( \pi^{ab}\mathcal{L}_\xi g_{ab})= \int d^3 x \left(\sigma^{ab}\mathcal{L}_\xi \rho_{ab}+\frac{\tau}{3}(\rho^{ab}\mathcal{L}_\xi\rho_{ab}+3\phi^{-1/3}\mathcal{L}_\xi\phi^{1/3}\right)\\
=\int d^3 x\left(\sigma^{ab}\mathcal{L}_\xi \rho_{ab}+\frac{\tau}{3}\mathcal{L}_\xi\ln{\phi} \right) =\int d^3 x\left(\sigma^{ab}\mathcal{L}_\xi \rho_{ab} \right) 
\end{multline}
where in the  third equality we used integration by parts and the fact that $\tau$ is a spatial constant. In the second equality, since $\rho_{ab}$ is a tensor density of weight $-2/3$, for the Lie derivative we have:
\be\label{equ:Lie_derivative} \mathcal{L}_\xi \rho_{ab}=\xi^c\rho_{ab|c}-\xi^c_{\phantom{c}|a}\rho_{cb}-\xi^c_{\phantom{c}|b}\rho_{ca}-\frac{2}{3}\xi^c_{\phantom{c}|c}\rho_{ab}
\ee where the solid vertical bar denotes covariant differentiation with respect to $\rho^{ab}$ (and thus $\xi^c\rho_{ab|c}=0$). The trace of expression \eqref{equ:Lie_derivative} with respect to  $\rho^{ab}$ vanishes. 

 Let the ADM action be given by the Legendre transform of the total Hamiltonian 
\be\label{equ:ADM_action} \mathcal{S}=\int dt \int _\Sigma d^3x (\pi^{ab}\dot g_{ab}-NS-\xi^c H_c)(x)
\ee
Ideally, after reduction  we would be able to rewrite \eqref{equ:ADM_action} as (ignoring the spatial diffeomorphisms for now):
\be\label{equ:red_Hamiltonian} \mathcal{S}=\int dt \int _\Sigma d^3x (\sigma^{ab}\dot \rho_{ab}-\mathcal{H}[\tau, \sigma^{ab},\rho_{ab};x))
\ee
thus being able to identify $\int d^3 x \mathcal{H}[\tau, \sigma^{ab},\rho_{ab};x)$ as the true Hamiltonian generating evolution in the reduced phase space. To see that this is possible, we must look only at a rewriting of the symplectic form $\int \pi^{ab}\dot g_{ab}$.

Using \eqref{equ:modified_PB} for calculating $\dot\rho_{ab}$ from the Poisson bracket with the Hamiltonian, it is trivial to see that  $\dot\rho_{ab}$ will contain a traceless projection. \footnote{ To be precise
$\dot\rho_{ab}(x):=\{\rho_{ab}(x), S(N)+H_i(\xi^i)\}= 2(\delta_{(a}^c\delta_{b)}^d-\frac{1}{3}g_{ab}g^{cd})(N\sigma_{cd}+\phi^{-5/3}\xi_{(c|d)})
$}
 In particular, we get that $\rho^{ab}\dot\rho_{ab}=0$. 
 Now,  from  the inverse transformation  \eqref{equ:inv_transf} we get
\be \int d^3 x  \pi^{ab}\dot g_{ab}=\int d^3 x \left(\frac{2}{3}\tau \dot{(\ln \phi)}+\sigma^{ab}\dot\rho_{ab}+\frac{\tau}{3}\rho^{ab}\dot\rho_{ab}\right)=
\int d^3x (\ln\phi+\sigma^{ab}\dot\rho_{ab})
\ee
where we have used $\rho^{ab}\dot\rho_{ab}=0$ and $\dot\tau=1$ and integration by parts. 
But as we have seen, the scalar constraint can be solved in full generality by 
a unique functional $\phi=F[\tau, \sigma^{ab},\rho_{ab};x)$.

Thus we can simultaneously do a phase space reduction by defining the variables $\phi:=F[\tau, \sigma^{ab},\rho_{ab};x)$ and by setting $\tau$ to be a spatial constant defining York time, i.e. $\dot \tau=1$. Of course, this incorporates the fact that the gauge-fixing $\tau-t=0$ is second class with respect to $S(x)=0$, and thus we must symplectically reduce to get rid of these constraints. 
We  are then  left with a genuine evolution Hamiltonian:
\be\label{equ:final_Ham} \mathcal{S}=\int dt \int _\Sigma d^3x (\sigma^{ab}\dot \rho_{ab}-\ln{\phi_o}-\sigma^{ab}\mathcal{L}_\xi \rho_{ab} )
\ee
note that the last term 
$ \sigma^{ab}\mathcal{L}_\xi \rho_{ab}$ now only generates diffeomorphisms whose flux is divergenceless (incompressible), since
$$\{\int d^3 x \sigma^{cd}\mathcal{L}_\xi \rho_{cd}, \rho_{ab}\}= (\delta_{(a}^c\delta_{b)}^d-\frac{1}{3}g_{ab}g^{cd})\mathcal{L}_\xi \rho_{cd}
$$ 

One of the drawbacks of using  a reference density $|\gamma|$ so that our variables have a specific form of coordinate-covariance,  is that the projected value of any functional $F[g,\pi]\rightarrow F[\rho, \sigma]$ is a priori dependent on the auxiliary (background) metric $\gamma_{ij}$. The only case where this is not so is if the functional $F[g,\pi]=F[\rho, \sigma]$ is already  conformally invariant.

This holds also if we had chosen not to introduce an auxiliary metric $\gamma_{ij}$, but worked with the original \eqref{equ:new_variables} instead. This can be seen as follows: although  the replacement $(g_{ij}\pi^{ij})\mapsto (\rho_{ij}, \sigma^{ij})$ is local, it  is also coordinate-dependent (due to the different density weights) and thus ambiguous.  One of the main advantages of Shape Dynamics is that one has diffeomorphism invariance intact for the full variables $g_{ab}, \pi^{ab}$ \emph{without the need to introduce auxiliary quantities}.

Another drawback of using the ADM in CMC  formalism is that  in order to reconstruct the metric (and all the usual physically meaningful quantities calculated with the full metric), one needs to reinsert the non-local York factor. It is the result of Shape Dynamics that you can have a different theory which also reduces to these variables, but which is expressed (fully locally) in terms of the physically meaningful full 3-metric.

\section{Machian Principles}\label{sec:Mach}

There are many inequivalent statements of Mach's principle and for each of these there many inequivalent mathematical implementations; we will therefore provide only aspects of Machian ideas that where relevant for the development of Shape Dynamics. Rather, we start with the prose postulate ``There is no absolute space or absolute time. Rather space and time are concepts that are abstracted from the relations of physical objects.'' This statement is open to numerous interpretations, so to make it precise we apply it to classical field theory with gravitational and matter degrees of freedom. We then consider the following aspects of relationalism:
\begin{enumerate}
 \item ``Equilocality is abstracted from the evolution of physical degrees of freedom.'' Using Barbour's idea of ``best matching,'' one can readily translate this statement into spatial diffeomorphism invaraince. The canonical formulation of a field theory should thus contain spatial diffeomorphism constraints.
 \item ``Spatial scale is abstracted from local ratios, i.e. ratios with physical rods.'' This can be readily translated into the requirement that the theory possess local spatial conformal invariance (in physicist terms: spatial Weyl-invariance), i.e. the canonical formulation possess local spatial conformal constraints.
 \item ``Time is abstracted from the dynamics of local physical degrees of freedom, i.e. the dynamics of physical clocks.'' We implement this by requiring that the theory posses local time reparametrization invariance. This means that the canonical formulation of the theory possesses local Hamilton constraints which generate infinitesimal spacetime refoliations.
\end{enumerate}
It is clear that GR implements spatial diffeomorphism- and refoliation-invariance, because the canonical formulation posses spatial diffeomorphism constraints and scalar constraints that generate on-shell refoliations. However, GR does {\bf not} possess local spatial conformal invariance. Rather, the fact that one can solve the scalar constraints  by using spatial conformal transformation, shows that the scalar constraints of GR gauge-fix conformal constraints. In other words: the generators of time reparametrizations in ADM - i.e. the scalar constraints -  and conformal constraints form a second class system.

Shape Dynamics on the other hand implements spatial diffeomorphism invariance and local spatial conformal invariance, but it posses a global Hamiltonian and thus fails to implement local time reparametrization invariance. This is of course expected from the equivalence with GR, since otherwise the number of local physical degrees of freedom would not be compatible. 

\section{Linking Theory and Symmetry Trading}\label{sec:LT-ST}

Physical observables of gauge theories are equivalence classes of gauge invariant phase space functions, where two phase space functions are equivalent if and only if their restrictions to the initial value surface coincide. This dual definition of observables is often necessary for a local description of a field theory, but it is also the reason why gauge symmetries can be traded. The simplest way to see how one gauge symmetry can be traded for another is through a {\it linking theory}. An instructive example linking theory is the following: Consider a dynamical system with elementary Poisson brackets $\{q^a,p_b\}=\delta^a_b$ for all $a,b \in \mathcal I$ and $\{\phi^\alpha,\pi_\beta\}=\delta^\alpha_\beta$ for all $\alpha,\beta \in \mathcal A$ and and all other elementary Poisson brackets vanishing. Assume the following first class set of constraints:
\begin{equation}\label{equ:special-LT-constraints}
 \chi_1^\alpha=\phi^\alpha+\phi_o^\alpha(q,p),\,\,\,\chi^2_\alpha=\pi_\alpha-\pi^o_\alpha(q,p),\,\,\,\chi_3^\mu=\chi_3^\mu(q,p),
\end{equation}
where $\alpha\in\mathcal A$ and $\mu$ in an index set $\mathcal M$ and where the Hamiltonian is contained in the set $\chi^3_0(q,p)=H(p,q)-E$ as an energy conservation constraint. There are two sets of interesting partial gauge-fixings for this system:
\begin{equation}
 \sigma^1_\alpha=\pi_\alpha\,\,\,\textrm{ and }\,\,\,\sigma^\alpha_2.
\end{equation}
Imposing the gauge fixing conditions $\sigma^1_\alpha$ leads to the phase space reduction $(\phi^\alpha,\pi_\beta)\to(-\phi^\alpha_o(q,p),0)$ where the Dirac bracket associated with this phase space reduction coincides with the Poisson bracket on the reduced phase space, because . I.e. the elementary Poisson brackets are $\{q^a,p_b\}=\delta^a_b$. The remaining first class constraints on reduced phase space are
\begin{equation}
  \chi^2_\alpha=-\pi^o_\alpha(q,p),\,\,\,\chi_3^\mu=\chi_3^\mu(q,p).
\end{equation}
Imposing on the other hand $\sigma^\alpha_2$ leads the the phase space reduction $(\phi^\alpha,\pi_\beta)\to(0,\pi^o_\alpha(q,p))$. The Dirac bracket again coincides with the Poisson bracket on reduced phase space and the remaining first class constraints are
\begin{equation} 
 \chi_1^\alpha=\phi_o^\alpha(q,p),\,\,\,\chi_3^\mu=\chi_3^\mu(q,p).
\end{equation}
The set of observables as well as their dynamics coincide for the two reductions, because they are obtained as partial gauge fixings of the same initial system. We call this initial system together with the two partial gauge fixing conditions as {\it linking theory}. The linking theory allows us to describe the same dynamical system with two different sets of first class constraints. These two sets of constraints do in general generate two different sets of gauge transformations. This means that linking theories enable us to trade one set of gauge symmetries for another.

Since partial gauge fixing and phase space reduction depend only on the constraint surface and not on the particular set of constraints, one sees that the constraints do not have to take the special form of equation (\ref{equ:special-LT-constraints}), but any equivalent form of these constraints is admissible for the definition of a linking theory. The only thing that is important for our construction is that the set of canonical pairs $(\phi^\alpha,\pi^\alpha)$ can be split into two sets of proper gauge fixing conditions $\phi^\alpha$ and $\pi_\alpha$. This freedom enables us to perform {\it symmetry trading}  between very general classes of gauge symmetries.

\subsection{Kretschmannization and Best Matching}\label{sec:BM}

Very often one is given a particular gauge theory and the goal is to simplify the gauge transformations by trading a complicated set of gauge transformations for a set of gauge transformations that closes on configurations space, i.e. transformations of the form $q^a \to Q^a(q,\phi)$, where $\phi^\alpha$ denote group parameters. A very useful tool for the construction of a linking theory that proves equivalence between the two systems is given by a canonical implementation of Barbour's ``best matching.'' One starts with the original system with first class constraints $\chi^\mu(q,p)\approx 0$ and extends phase space by the cotangent bundle over the gauge group. To make this extension pure gauge, one introduces additional first class constraints that require the momenta conjugate to the group parameters to vanish $\pi_\alpha\approx 0$. Then one employs a canonical transform generated by \begin{equation}
 F=P_aQ^a(q,\phi)+\Pi_\alpha\phi^\alpha,
\end{equation}
which generates a canonical transformation that takes $q^a \to Q^a(q,\phi),$ $p_a\to (Q_{,q}^{-1})^b_a p_b,$ $\phi^\alpha\to \phi^\alpha$ and takes the additional constraints to
\begin{equation}\label{equ:transformed-constraints}
 \pi_\alpha \to \pi_\alpha - Q^a_{,\alpha}(Q_{,q}^{-1})^b_ap_b,
\end{equation}
where $(Q_{,q}^{-1})$ denotes the inverse of the matrix $\partial_{q^b}Q^a(q,\phi)$. We have now {\it Kretschmannized} the system, i.e. we have implemented a trivial gauge symmetry in by enlarging the phase space. Canonical {\it best matching} is achieved by requiring that the group transformations are pure gauge, i.e. by imposing the conditions
\begin{equation}
  \pi_\alpha\approx 0.
\end{equation}
Performing a Dirac analysis for the best matching conditions leaves many possibilities; the most interesting in light of the previous subsection is that a subset of the constraints $\chi^\mu(Q(q,\phi),Q_{,q}^{-1}p)$ can be solved for $\phi^\alpha$. This means that this subset of the constraints can be equivalently expressed as $\phi^\alpha-\phi^\alpha_o(q,p)\approx 0$. Moreover, if a group parametrization is chosen s.t. $\phi^\alpha\equiv0$ denotes the unit element, then $\phi^\alpha_o(q,p)\approx 0$ is equivalent to imposing the original subset of constraints. This means that we have constructed a linking theory through best matching, because imposing $\phi^\alpha\equiv 0$ gauge-fixes the constraints of equ (\ref{equ:transformed-constraints}) and the remaining constraints describe the original system. Imposing the best matching condition $\pi_\alpha\equiv0$ on the other hand turns gauge fixes the constraints $\phi^\alpha_o(q,p)$ and allows us to trade them for $Q^a_{,\alpha}(Q_{,q}^{-1})^b_ap_b$, which are 
linear in the momenta and generate the transformations $q^a \to Q^a(q,\phi)$ on configuration space. Canonical best matching is thus a way to construct a linking gauge theory, whenever a subset of the constraints $\chi^\mu(Q(q,\phi),Q_{,q}^{-1}p)$ can be solved for $\phi^\alpha$.

\section{Poincar\'e invariance in the flat solution}\label{sec:Poincare}

Here we briefly describe how a solution corresponding to Minkowski spacetime has  Poincar\'e Symmetry. Full conformal symmetry is excised for the standard choice of boundary conditions, but might emerge for a different choice.

In this particular case we will be dealing with the curve of phase space data $(g_{ab}(t), \pi^{ab}(t))=(\delta_{ab}, 0)$, which thus is already in maximal slicing. One can easily see that the LY equation is simply written as $\partial^2 \Omega=0$, where $\partial^2$ is the Laplacian for $\delta_{ab}$. In rectilinear coordinates $\{x^a\}$ the set of solutions is given by $\{1, x^a\}$. The lapse fixing equation in the Linking Theory is given by:
\be\label{equ:tphi_LFE}e^{-4\phi}(\nabla^2 N+2g^{ab}\phi_{,a}N_{,b}) -Ne^{-6\phi} G_{abcd}\pi^{ab}\pi^{cd}=0
\ee
Over our set of data the solutions to this equation can be divided into two sets, one for $\Omega=1$, and one when $\Omega=x^a$. When $\Omega=1$ the solutions are given by $N_o^{(i)}=\{1, x^a\}$.

The Hamiltonian for Shape Dynamics is (naively) given by 
\be \mathcal{H}(N^{(i)}_o, \xi^a, \rho)=t_{\phi_o}S(N^{(i)}_o)+H_a(\xi^a)+\pi(\rho)
\ee
where only the solutions of the lapse fixing equation are allowed in the smearing. The algebra of constraints emerging from this (ignoring boundary terms)  is
\begin{multline}\label{equ:SD_algebra} [\mathcal{H}(N_o^{(i)}, \xi^a, \rho), \mathcal{H}(N_o^{(i')}, \xi'^{a'}, \rho')]=\\
\mathcal{H}\left((\xi'^{a'}{N^{(i)}_o}_{,a'}-\xi^{a}{N^{(i')}_o}_{,a}), ~(g^{cd}({N^{(i)}_o}_{,c}N_o^{(i')}-{N^{(i')}_o}_{,c}N^{(i)})+ [\mathbf{\xi},\xi']^d),~ \xi'^{a'}\rho_{,a'}-\xi^{a}\rho'_{,a}\right)
\end{multline}
 Using the Linking Theory equations of motion, one can check  that for  $\Omega=1$  the following smearings generate symmetries of the data (i.e $\dot g_{ab}=\dot \pi^{ab}=0$): time translations and boosts are given respectively by $N_o^{(i)}={1, x^a}$, translations along the $c$ coordinate $\xi^a=\delta^a_{(c)}$ and rotations around the $d$ axis $ \xi^a=\epsilon_{ad(c)}x^d$. Using the algebra \eqref{equ:SD_algebra} one checks that indeed these reproduce the usual Poincar\'e algebra. This is clarified in \cite{Gomes:Poincaré}

\end{appendix}

\end{document}